\documentstyle[twocolumn]{article}

\def\ZzZ{{\hbox{\tenrm Z\kern-.31em{Z}}}} 
\def\CcC{{\hbox{\tenrm C\kern-.45em{\vrule height.67em width0.08em depth- 
.04em 
\hskip.45em }}}}

\newcommand{\bc}{\begin{center}} 
\newcommand{\ec}{\end{center}} 
\newcommand{\be}{\begin{equation}} 
\newcommand{\ee}{\end{equation}} 
\newcommand{\bea}{\begin{eqnarray}} 
\newcommand{\eea}{\end{eqnarray}} 
\newcommand{\bs}{\begin{subequations}} 
\newcommand{\es}{\end{subequations}} 
\newcommand{\beq}{\begin{eqalignno}} 
\newcommand{\eeq}{\end{eqalignno}} 

\begin{document}

\thispagestyle{empty} 
 
\bc 

{\Large{ \bf Life-time and hierarchy of memory in 
the dissipative quantum model of brain}}

\vspace{8mm}

\large{Eleonora Alfinito and Giuseppe Vitiello} \\ 
\small 

{\it Dipartimento di Fisica, Universit\`a di Salerno} \\ 
{\it 84100 Salerno, Italia and INFM Unit\`a di Salerno} \\
\indent {\it alfinito@physics.unisa.it}\\
\indent {\it vitiello@physics.unisa.it }\\ 
\vspace{1.3cm}

\ec 
\small 

\normalsize



%
\newcommand{\bib}{\bibitem}









In this report we will consider some recent developments of the quantum
model of brain\cite{UR,S1,S2,CH} which include dissipative
dynamics\cite{VT} and time dependent frequency of the electric dipole
wave quanta (dwq)\cite{AV}. The dissipative quantum model of brain has
been recently investigated\cite{PV} also in relation with the
possibility of modeling neural networks exhibiting collective dynamics
and long range correlations among the net units. The study of such a
quantum dynamical features for neural nets is of course of great
interest either in connection with computational neuroscience, either in
connection with quantum computational strategies based on quantum
evolution (quantum computation). On the other hand, further developments
of the quantum model of brain, on which here we do not report, show
attractive features also related with the r\^ole of microtubules in
brain activity\cite{PR,P2,YA}.

In previous works\cite{UR,S1,S2,VT} it has been considered the case  of
time independent frequencies  associated to each dwq. A more general
case is the one where time dependent frequencies are also
considered. Such a case is of course more appropriate to realistic
situations where the dwq may undergo a number of fluctuating
interactions with other quanta and then their characteristic frequency
may  accordingly change in time. The study of the dissipative model with
time dependent frequency leads to a number of interesting new features
some of which we will briefly discuss in the following.

Let us first summarize few aspects of the quantum brain   model. We
remind that the observable   specifying the ordered state is called the
order parameter and acts   as a macroscopic variable for the system. The
order parameter is   specific of the kind of symmetry into play and may
thus be   considered as a code specifying the vacuum or ground state of
the   system. The value of the order parameter is related with the
density of condensed Goldstone bosons in the vacuum and   specifies the
phase of the system with relation to the   considered symmetry. Since
physical properties are different for   different phases, also the value
of the order parameter may be   considered as a code number specifying
the system state.

In the quantum model of brain the  information  storage    function  is
represented  by  the  coding  of  the  ground  state   through the
condensation of the dipole wave quanta.

Suppose a vacuum of specific code number has been selected by  the
printing of a specific information. The brain then sets in   that state
and no other vacuum state is successively accessible   for recording
another information, unless   the external stimulus carrying the new
information   produces a phase transition to   the vacuum specified by
the new code number.   This will destroy the previously stored
information, we have  {\it overprinting}: vacua labeled by different
code numbers are  accessible only through a sequence of phase
transitions from one to   another one of them.
 
It can be shown\cite{VT} that by taking into account the   fact that the
brain is an open system one may reach a  solution to the problem of
memory capacity:  infinitely many vacua are accessible to memory
printing in such   a way that in a sequential information recording the
successive   printing of information does not destroy the previously
recorded ones; a huge memory capacity is thus achieved.
 
In the quantum brain model spontaneous breakdown of dipole   rotational
symmetry is triggered by the coupling of the  brain with external
stimuli. Let us remark  that {\it once} the dipole rotational   symmetry
has been broken (and information has thus been   recorded), {\it then,
as a consequence}, time-reversal  symmetry is also broken: {\it Before}
the information recording   process, the brain can in principle be in
anyone of  the infinitely many (unitarily inequivalent) vacua.  {\it
After} information has been recorded, the brain state is   completely
determined and the brain cannot be brought to   the state configuration
in which it was {\it before} the  information printing occurred.  Thus,
the same fact of   getting information introduces {\it the arrow of
time}   into brain dynamics. Due to memory printing process time
evolution of the brain states is intrinsically irreversible.  Getting
information introduces a partition   in the time evolution, it
introduces the {\it distinction}   between the past and the future, a
distinction which did not exist  {\it before} the information recording.

Let us now illustrate in more details how the quantum dissipation
formalism\cite{VT} allows to solve the overprinting problem in  the
quantum model of brain.
 
The mathematical treatment of quantum dissipation requires the
"doubling"of the degrees of freedom of the dissipative system.  Let
$A_{\kappa}$  and ${\tilde A}_{\kappa}$ denote the dwq  mode and its
"doubled mode", respectively. The  $\tilde A$ mode is the "time-reversed
mirror image"  of the $A$ mode and represents the environment mode.  Let
${\cal N}_{A_{\kappa}}$ and ${\cal N}_{{\tilde A}_{\kappa}}$  denote the
number of ${A_{\kappa}}$  modes and ${\tilde A}_{\kappa}$ modes,
respectively.
 
Taking into account dissipativity requires that  the memory state,
identified with the vacuum ${|0>}_{\cal N}$ ,  is a condensate of {\it
equal number} of $A_{\kappa}$  and ${\tilde A}_{\kappa}$ modes, for any
$\kappa$ : such a requirement ensures in fact that the flow of the
energy exchanged between the system and the environment is
balanced. Thus, the difference between the number of tilde and non-tilde
modes must be zero:  ${\cal N}_{A_{\kappa}} - {\cal N}_{{\tilde
A}_{\kappa}} = 0$, for any $\kappa$.  Note that the label ${\cal N}$ in
the vacuum symbol ${|0>}_{\cal N}$  specifies  the set of integers
$\{{\cal N}_{A_{\kappa}}, ~for~any~  \kappa \}$ which indeed defines the
"initial value" of the condensate,  namely the {\it code} number
associated to  the information recorded at time $t_{0} = 0$. Note now
that the requirement  ${\cal N}_{A_{\kappa}} - {\cal N}_{{\tilde
A}_{\kappa}} = 0, $ for any $ \kappa$,  does not uniquely fix the set
$\{{\cal N}_{A_{\kappa}}, ~for~any~  \kappa \}$. Also  ${|0>}_{\cal N'}$
with ${\cal N'} \equiv \{ {\cal N'}_{A_{\kappa}};  {\cal
N'}_{A_{\kappa}} - {\cal N'}_{{\tilde A}_{\kappa}} = 0,  ~for~any~
\kappa \}$ ensures the energy flow balance and therefore also
${|0>}_{\cal N'}$ is an available memory state: it will correspond
however to a different code number  $(i.e. {\cal N'})$  and therefore to
a different information than the one of code ${\cal N}$. Thus,
infinitely many memory (vacuum) states, each one  of them corresponding
to a different  code $\cal N$, may exist: A huge number of sequentially
recorded informations may  {\it coexist} without   destructive
interference since infinitely many vacua  ${|0>}_{\cal N}$, for all
$\cal N$,  are {\it independently} accessible   in the sequential
recording process.  The "brain  (ground) state" may be represented as
the collection (or the  superposition) of the full set of memory states
${|0>}_{\cal N}$, for all  $\cal N$.
 
In summary, one may think of the brain as a complex   system with a huge
number of   macroscopic states (the memory states).   The degeneracy
among the vacua    ${|0>}_{\cal N}$ plays a crucial r\^ole in solving
the problem of memory capacity.  The  dissipative dynamics introduces
$\cal N$-coded "replicas" of the system   and information printing can
be  performed in each replica without destructive interference with
previously recorded informations in other replicas. In the
nondissipative   case the "$\cal N$-freedom" is missing and consecutive
information printing produces overprinting.
 
We remind that it does not exist in the infinite  volume limit any
unitary transformation which may transform one   vacuum  of code ${\cal
N}$ into another one of code ${\cal N'}$: this fact, which is a typical
feature of QFT, guarantees that  the corresponding printed informations
are indeed  {\it different} or {\it distinguishable} informations (
$\cal N$ is a  {\it good} code) and that each information printing is
also {\it protected}   against interference from other information
printing (absence of {\it   confusion} among  informations). The effect
of finite (realistic) size of the system may   however spoil  unitary
inequivalence. In the case of open systems, in fact,  transitions among
(would be) unitary inequivalent vacua may  occur (phase transitions) for
large  but  finite  volume,  due  to  coupling  with  external
environment.  The inclusion  of  dissipation  leads  thus  to  a picture
of the system "living over many  ground  states" (continuously
undergoing phase transitions).  It is to be noticed that even very weak
(although above a certain threshold) perturbations may drive the system
through its macroscopic configurations.  In this way, occasional
(random)  weak perturbations are recognized to play an important r\^ole
in the   complex behavior of the brain activity. The possibility of
transitions among different vacua is a feature of the model which is not
completely negative: smoothing out the exact unitary inequivalence among
memory states has the advantage of allowing the familiar phenomenon of
the "association" of memories:   once transitions among different memory
states are "slightly"  allowed the possibility of associations
("following a path of memories") becomes possible. Of course, these
"transitions" should only be allowed up to a certain degree in order to
avoid memory "confusion" and difficulties in the   process of storing
"distinct" informations.
 
According to the original  quantum brain model\cite{UR},  the recall
process is described as the excitation of  dwq modes  under an external
stimulus which is  "essentially a replication signal" of  the one
responsible for   memory printing. When dwq are excited   the brain
"consciously feels"  the presence of the   condensate pattern in the
corresponding coded vacuum.  The replication signal thus acts as a probe
by which the brain  "reads" the printed information. Since the
replication signal   is represented in terms of ${\tilde
A}$-modes\cite{VT} these modes act   in such a reading as the "address"
of the information to be recalled.  In this connection, we also observe
that  the dwq may acquire an effective non-zero mass due to the effects
of   the system finite size. Such an effective mass will then  act as a
threshold in the excitation energy of dwq so that,  in order to trigger
the recall process  an energy supply equal or greater than such a
threshold is required. When the energy supply is lower than the required
threshold a "difficulty in   recalling" may be experienced.  At the same
time, however, the   threshold may positively act as a "protection"
against unwanted   perturbations (including thermalization) and
cooperate to the stability of the memory  state.  In the case of zero
threshold any replication signal could excite  the recalling  and the
brain would fall in a state of "continuous flow of   memories".
 
We observe that we are considering memory states   associated to the
ground states and therefore of long life-time.   These states have a
finite (although long) life-time because   of dissipativity. Then, at
some time $t = \tau$,   conveniently larger than the memory life-time,
the memory state   $|0>_{\cal N}$ is reduced to the   "empty" vacuum
$|0>_{0}$ where  ${\cal N}_{\kappa} = 0$ for all $\kappa$:  the
information has been {\it forgotten}.  At the time $t = \tau$  the state
$|0>_{0}$ is available for recording a new information.  It is
interesting to observe that in order to not completely forget  certain
information,  one needs  to "restore" the ${\cal N}$ code, namely to
"refresh" the memory by   {\it brushing up} the subject (external
stimuli maintained memory). We thus see how another familiar feature of
memory can find a possible explanation in the dissipative quantum model
of brain.

It can be shown that the evolution  of the memory state   is controlled
by the entropy variations:   this feature indeed reflects the
irreversibility of time evolution (breakdown of time-reversal symmetry)
characteristic of dissipative systems, namely  the choice of a
privileged  direction in time evolution (arrow of time).  Moreover, the
stationary condition  for the  free energy functional leads to recognize
the memory state  $|0(t)>_{\cal N}$ to be a  finite temperature
state\cite{U2}. The dissipative quantum brain model thus also brings
us to the possibility of thermodynamic considerations in the brain
activity.

Until now we have considered the case of time independent dwq
frequencies.  In the case of time dependent frequencies one
finds\cite{AV} that dwq of different momentum $k$ acquire different
life-time values. Modes with longer life-time are the ones with higher
momentum. Since the momentum goes as the reciprocal of the distance over
which the mode can propagate, this means that modes with shorter range
of propagation will survive longer. On the contrary, modes with longer
range of propagation will decay sooner. The scenario becomes then
particularly interesting since this mechanism may produce the formation
of ordered domains of finite different sizes with different degree of
stability: smaller domains would be the more stable ones. Remember now
that the regions over which the dwq propagate are the domains where
ordering (i.e. symmetry breakdown) is produced. Thus we arrive at the
dynamic formation of a hierarchy (according to their life-time or
equivalently to their sizes) of ordered domains. Since in our case
"ordering" corresponds to the recording process, we have that the
recording of specific information related to dwq of specific momentum
$k$ may be "localized" in a domain of size proportional to $1/k$, and
thus we also have a dynamically generated hierarchy of memories. This
might fit some neurophysiological observations by which some specific
memories seem to belong to certain regions of the brain and some other
memories seem to have more diffused localization.  We thus see how the
dissipative quantum dynamics leads to a dynamic organization of the
memories in space (i.e. in their domain of localization) and in time
(i.e. in their persistence or life-time).
 
One more remark has to do with the "competition" between the frequency
term and the dissipative term in the dwq equation.  Such a competition
(i.e. which term dominates over the other one) may result in a smoothing
out of the dissipative term,  which may become then even  negligible,
or, on the contrary, in the enhancement of its dynamical r\^ole. In the
former case, according to the discussion in the present chapter, we
should expect a lowering of the memory capacity which could manifest in
a sensible "confusion" of memories, a difficulty in memorizing, a
difficulty in recalling, namely those "pathologies" arising from the
lost of the many degenerate and inequivalent vacua due to the lost of
dissipativity. In the latter case, the memory capacity would be
maintained high, but the hierarchy of memory life-time would result in a
strong inhibition of small momenta, i.e. the number of smaller domains
would be greater than the number of larger ones. Since smaller domains
have longer life-time, we see that more persistent memories would be
favored with respect to short-range memories. Which in some
circumstances also corresponds to
a commonly experienced phenomenon. To such a situation
would correspond an higher degree of localization of memories, but also
more sensible finite volume effects corresponding to effective non-zero
mass for the dwq and therefore again difficulties in recalling.
 
The competition or the balance between the frequency term and the
dissipative term in the dwq equation is controlled by parameters whose
values cannot be fixed by the model dynamics. They have to be given by
some external input of biochemical nature. It is quite interesting that
the dissipative quantum model contains such freedom in setting the
values for such parameters. We recall that Takahashi, Stuart and Umezawa
did suggested that the formation of ordered domains could play a
significant r\^ole in establishing the bridge between the basic dynamics
and the biomolecular phenomenology. Work along this direction is in
progress.



\begin{thebibliography}{9999}


\bibitem{UR}  L.M.Ricciardi and H.Umezawa, Brain physics 
and many-body
problems. {\it Kibernetik} {\bf 4}, 44 (1967)
\bibitem{S1} 
C.I.J.Stuart, Y.Takahashi and  H.Umezawa, 
On the stability and 
non-local 
properties of memory. {\it J. Theor. Biol.} {\bf 71}, 605 (1978)
\bibitem{S2} C.I.J.Stuart, Y.Takahashi and 
H. Umezawa, Mixed system brain dynamics: neural memory as a macroscopic
ordered state, {\it Found. Phys.} {\bf 9}, 301 (1979)    
\bibitem{CH} S. Sivakami and V. Srinivasan, 
A model for memory,
{\it J. Theor. Biol.} {\bf 102}, 287 (1983)
\bibitem{VT} G. Vitiello, Dissipation and memory capacity in the quantum
brain model. {\it Int. J. Mod. Phys.} {\bf 9} 973 (1995)
\bibitem{AV} E. Alfinito and G. Vitiello, in preparation
\bibitem{PV} E.Pessa and G.Vitiello, Quantum dissipation and neural 
net dynamics, {\it Biolectrochemistry and bioenergetics} {\bf 48}, 339 (1999)
\bibitem{PR} K.H.Pribram, {\it Languages of the brain}, 
Englewood Cliffs,  
New Jersey, 1971
\bibitem{P2} K.H.Pribram, {\it Brain and perception},   Lawrence
Erlbaum,  New  Jersey, 1991
\bibitem{YA} M.Jibu , K.H.Pribram  
and K.Yasue, From conscious experience to memory storage and retrivial:
The role of quantum brain dynamics and boson condensation of evanescent
photons. {\it Int. J. Mod. Phys.} {\bf B10}, 1735 (1996)   
\bibitem{U2} H.Umezawa, 
{\it Advanced field theory: micro, macro and thermal concepts}, 
American Institute of Physics, N.Y. 1993
 
\end{thebibliography}
\end{document}